\pdfoutput=1
\documentclass[reprint,showpacs,preprintnumbers,amsmath,amssymb,aps,nofootinbib]{revtex4-1}
\usepackage{ulem}
\usepackage{graphicx}
\usepackage[mathscr]{euscript}
\usepackage{dcolumn}
\usepackage[colorlinks=true, linkcolor = red, citecolor = blue]{hyperref}
\usepackage{subcaption}
\usepackage{natbib}
\usepackage{float}
\usepackage{comment}
\usepackage{braket}
\usepackage{xcolor}

\begin{document}
\title{Primordial Black Hole Formation in a Scalar Field Dominated Universe: Investigation of the Critical nature of the Collapse}

\author{Luis E. Padilla}
  \email{l.padilla@qmul.ac.uk}
\affiliation{Astronomy Unit, Queen Mary University of London, Mile End Road, London, E1 4NS, UK.}
\author{Ethan Milligan}
\email{e.milligan@qmul.ac.uk}
\affiliation{Astronomy Unit, Queen Mary University of London, Mile End Road, London, E1 4NS, UK.}
  \author{David J. Mulryne}
  \email{d.mulryne@qmul.ac.uk}
  \affiliation{Astronomy Unit, Queen Mary University of London, Mile End Road, London, E1 4NS, UK.}
    \author{Juan Carlos Hidalgo}
  \email{hidalgo@icf.unam.mx}
  \affiliation{Instituto de Ciencias Físicas, Universidad Nacional Autónoma de México, 62210, Cuernavaca, Morelos, México.}
  
\date{\today}%

\begin{abstract}
In this paper, we investigate the critical collapse leading to primordial black hole (PBH) formation in a universe dominated by a self-interacting scalar field with a quartic potential, comparing it to the well-known radiation-dominated case. Using fully relativistic nonlinear numerical simulations in spherical symmetry, based on the Misner--Sharp formalism, we analyze the dynamics near the collapse threshold and track the scaling of the black hole mass. Our results confirm that both the scalar field and radiation cases exhibit type II critical behavior with similar --- though not identical --- critical exponents, differing by about \( 2\sigma \). This suggests that, while a quartic scalar field effectively mimics a radiation fluid even in the nonlinear collapse regime, small differences in the critical exponent persist. Our findings provide direct numerical evidence for the near universality of the critical exponent in PBH formation, with only mild dependence on whether the collapse is driven by a scalar field or a perfect fluid.

\end{abstract}

\maketitle

\section{Introduction}

Gravitational collapse near the threshold of black hole formation exhibits striking parallels with phase transitions in statistical physics. When tuning a parameter \( p \) in the initial data of a collapsing system, one often finds a critical value $ p_{\rm th}$ separating two distinct outcomes: for $p < p_{\rm th}$, the spacetime remains globally regular, while for $p > p_{\rm th}$, a black hole forms. Near this threshold, the mass of the resulting black hole follows a characteristic power-law scaling of the form
\begin{equation}
    M \propto (p - p_{\rm th})^\gamma,
\end{equation}
where \( \gamma \) is a critical exponent. This phenomenon, known as critical gravitational collapse, was first discovered numerically by Choptuik \citep{PhysRevLett.70.9} in the case of a real massless scalar field and has since been observed across a variety of matter fields and environments, including radiation fluids \citep{PhysRevD.59.124013, PhysRevLett.80.5481,IHawke_2002,Musco_2005,Polnarev_2007,Musco:2008hv}, Yang--Mills fields \citep{PhysRevLett.77.424}, and massive scalar fields \citep{PhysRevD.56.R6057,PhysRevD.62.104024,Jimenez-Vazquez:2022fix}.

Critical collapse is typically classified into two types \citep{Gundlach:2007gc}: type I, where the black hole mass has a finite minimum value and the critical solution is either stationary or periodic; and type II, where the mass can become arbitrarily small and the critical solution displays discrete or continuous self-similarity. The presence of an intrinsic mass or length scale in the field equations often determines the type of critical behavior. For instance, systems governed by scale-free dynamics, such as a massless scalar field or a perfect fluid with radiation equation of state, typically exhibit type II collapse. In contrast, massive or self-interacting fields may display type I and type II behavior depending on how the scale associated with the potential compares with the initial data profile.

When dealing with scalar fields, a common practice in the literature on primordial black hole (PBH) formation is to assume an effective equivalence with perfect fluids in certain regimes (see, for example, \citep{Hidalgo:2017dfp, Ralegankar:2024zjd, Carrion:2021yeh, Martin:2019nuw,Brady:2002iz,Harada:2004pf, Padilla:2021uof}). This simplification is largely motivated by the fact that, in the limit of rapid oscillations, a scalar field with a polynomial potential, 
\begin{equation}
    V(\psi) = \frac{\lambda_{2n}}{2n} \psi^{2n},
\end{equation}
behaves as a fluid with an effective equation of state 
\begin{equation}
    \omega = \frac{n - 1}{n + 1},
\end{equation}
as discussed in~\citep{PhysRevD.28.1243}. For instance, a quadratic or a quartic self-interacting scalar field, with $n = 1$ or $n = 2$, is often modeled as a matter- or radiation-like fluid in an averaged sense. This correspondence has justified the adoption of numerical results for perfect fluids in studies of scalar field systems and PBH formation. However, this analogy holds strictly only when the scalar field oscillates rapidly around the minimum of its potential---a condition that may break down in strongly nonlinear regimes such as gravitational collapse. 

In a previous work \citep{Milligan:2025zbu}, we presented numerical simulations of PBH formation from a universe dominateed by quadratic and quartic cases. In our initial results, we determined the collapse threshold for the quartic potential, finding it to be very similar to that of PBH formation during radiation domination. By contrast, in the quadratic case we observed clear departures from the dust analogy. In particular, we found a flattening of the central density of the evolving profiles—a feature that reflects the wave-like nature of the scalar field and is consistent with the repulsive effect of the scalar field gradients (or quantum potential) associated with the system, as suggested in Refs.~\citep{Padilla:2021zgm,Hidalgo:2022yed,Padilla:2023lbv,Padilla:2024cbq,Niemeyer:2019gab,Padilla:2024iyr}.

Building on our previous work in scalar field collapse, our goal in this paper is to test how far the perfect-fluid analogy holds when approaching the regime of critical collapse. We have already seen that discrepancies with the fluid description appear for scalar fields with a quadratic potential—where even the early evolution of perturbations deviates from dust-like behavior, and the detection of apparent horizons is generally easier in the scalar field case than in dust. However, a detailed comparison between scalar fields with other potentials and perfect fluids in the nonlinear regime remains largely unexplored. In particular, it is still unclear whether a quartic self-interacting scalar field exhibits the same scaling laws and critical exponents as a radiation fluid.

To address this question, we perform fully nonlinear numerical simulations of critical collapse for a scalar field with a quartic potential and compare the results with those of a standard radiation fluid. We find that both systems display remarkably similar critical behavior, including matching critical exponents within numerical accuracy. This provides the first explicit verification (to our knowledge) that the scalar field–fluid correspondence extends to the critical collapse regime—at least for the quartic case—and supports the common practice of modeling such scalar field scenarios using perfect fluids.

At the same time, our study highlights the importance of explicitly verifying such assumptions, especially in applications to PBH formation in the early universe. The dynamics of collaps and the resulting PBH mass spectrum depend sensitively on the matter model, and seemingly minor differences in the collapse behavior can have significant phenomenological implications.

The paper is organized as follows. In Section~\ref{Sec:II}, we present the main field equations for the quartic self-interacting scalar field and the radiation fluid, along with the initial data used in our simulations. Section~\ref{Sec:III} describes the criteria for PBH formation and the method used to compute the final black hole mass. In Section~\ref{Sec:IV}, we discuss the numerical methods and consistency checks we employed. Section~\ref{Sec:V} presents our main results, including the determination of critical exponents and the characterization of the critical behavior in the radiation- and scalar field-dominated scenarios. Finally, in Section~\ref{Sec:VII}, we present some final comments and provide our conclusions.

\section{Mathematical formulation of the problem}\label{Sec:II}

In this section, we outline the formalism and initial conditions used to numerically track the evolution of perturbations in both radiation- and scalar field-dominated universes. While we present all quantities in a general form, we will later specialize to the simpler cases where the matter content consists solely of either a radiation fluid or a scalar field in a quartic potential.

\subsection{The Misner-Sharp formalism}

In this work, we adopt the Misner–Sharp formalism, commonly used to study the collapse of spherically symmetric systems \citep{PhysRev.136.B571} (see also Refs.~\citep{Niemeyer:1999ak,Escriva:2019nsa, Musco:2004ak, Nakama:2013ica, Bloomfield:2015ila} for some examples). In this formalism, the spacetime line element is given by
\begin{equation}\label{eq:MSmetric}
ds^2 = -e^{2\phi}dt^2 + e^{\lambda}dA^2 + R^2d\Omega^2,
\end{equation}  
where the metric functions $\phi$, $\lambda$, and the areal radius $R$ depend solely on the radial coordinate $A$ and time $t$.

For the matter sector, we include a perfect fluid and a scalar field $\psi$. The presence of the perfect fluid is essential, as it defines the comoving frame required by the Misner–Sharp formalism. The combined energy–momentum tensor takes the form:
\begin{eqnarray}
    T_{\mu\nu} &=& (\rho_{\rm pf} + P_{\rm pf})u_{\mu}u_{\nu} + P_{\rm pf} g_{\mu\nu} \nonumber \\
    && - g_{\mu\nu}\left(\frac{1}{2}\partial_{\sigma}\psi\partial^{\sigma}\psi + V(\psi)\right) + \partial_{\mu}\psi\partial_{\nu}\psi.
\end{eqnarray}
Then, the comoving frame of the perfect fluid can be expressed as  
\begin{equation}\label{eq:comovil}
u^t = e^{-\phi}, \quad u^i = 0 \quad \text{for } i = r, \theta, \phi.
\end{equation}

To facilitate the analysis, we introduce auxiliary variables:  
\begin{eqnarray}\label{eq:ugam}
U = e^{-\phi}\dot{R}, &\quad& \Gamma = e^{-\lambda/2}R',\nonumber \\
\chi \equiv \psi^{\prime}, &\quad& \Pi \equiv \dot{\psi}e^{-\phi}.
\end{eqnarray}  
Here, \( U \) represents the physical radial velocity of the fluid as seen by an Eulerian observer, and \( \Gamma \) generalizes the Lorentz factor in this geometry.

The evolution of the scalar field + a perfect fluid in spherical symmetry is governed by the following set of differential equations:
\begin{subequations}\label{eq:MS}
\begin{equation}
    \dot\psi = \Pi e^\phi,
\end{equation}
\begin{equation}
    \dot\chi = (\Pi^{'}+\phi^{'}\Pi )e^\phi,
\end{equation}
\begin{eqnarray}
    \dot\Pi &=& e^\phi\left\lbrace e^{-\lambda}\left[\left(\frac{2R'}{R} + \phi' - \frac{\lambda'}{2}\right)\chi + \chi'\right]\right. \\
    && -\left.\left[\left(\frac{2U}{R}+\frac{4\pi R\Pi\chi + U^{'}}{R'}\right)\Pi + \frac{dV}{d\psi}\right]\right\rbrace,\nonumber
\end{eqnarray}
\begin{equation}\label{eq:Rdot}
    \dot{R} = U e^{\phi},
\end{equation}
\begin{equation}\label{eq:mpfdot}
    \dot \rho_{\rm pf} = -e^{\phi}\left(\frac{2U}{R}+\frac{4\pi R \Pi\chi + U^{'}}{R'}\right)(\rho_{\rm pf} + P_{\rm pf}),
\end{equation}
\vspace{-0.25cm}
\begin{equation}\label{Udot}
\dot{U} = -e^{\phi}\left(\frac{e^{-\lambda}R'P_{\rm pf}'}{\rho_{\rm pf} + P_{\rm pf}} + \frac{8\pi P_T R^2 + 1 + U^2 - \Gamma^2}{2R}\right),
\end{equation}
\begin{equation}
    \dot\lambda = \frac{8\pi R\Pi\chi + 2U^{'}}{R'}e^{\phi},
\end{equation}
\begin{equation}\label{eq:rhopf}
    \phi^{\prime} = -\frac{P_{\rm pf}^{\prime}}{P_{\rm pf} + \rho_{\rm pf}},
\end{equation}
\begin{equation}\label{eq:rhosf}
    \rho_{\rm pf} = \frac{m_{\rm pf}^{\prime}}{4\pi R^2 R^{\prime}},
\end{equation}
\begin{equation}
    \rho_\psi = \frac{m_\psi^{\prime}}{4\pi R^2 R^{\prime}},
\end{equation}
\begin{equation}\label{eq:Gam1}
    \Gamma^2 = 1 + U^2 - \frac{2m_{\rm T}}{R} + \frac{2I}{R}.
\end{equation}
where
\begin{equation}\nonumber
    I \equiv 4\pi \int^A_0 R^2 U \chi\Pi dA.
\end{equation}
\end{subequations}
In the above equations, derivatives with respect to the radial coordinate $A$ are denoted by a prime while a dot indicates derivation with respect to the cosmic time $t$. The functions $m_{\rm pf}$, $m_\psi$, and $m_{\rm T} = m_{\rm pf} + m_\psi$ denote the local mass contributions from the fluid, scalar field, and total mass, respectively. The energy density and pressure associated with the scalar field are given by:
\begin{subequations}\label{eq:tilde}
\begin{equation}\label{eq:sfed}
    \rho_\psi = \frac{e^{-2\phi}}{2}\dot{\psi}^2 + \frac{e^{-\lambda}}{2}(\psi')^2 + V(\psi),
\end{equation}
\begin{equation}
    P_\psi = \frac{e^{-2\phi}}{2}\dot{\psi}^2 + \frac{e^{-\lambda}}{2}(\psi')^2 - V(\psi).
\end{equation}
\end{subequations}
The total pressure in the system is then \( P_T = P_{\rm pf} + P_\psi \).

To ensure regularity at the origin, we impose boundary conditions such that \( R(t,0) = 0 \) and \( U(t,0) = 0 \). Furthermore, spherical symmetry requires all first derivatives with respect to \( A \) to vanish at \( A = 0 \), i.e.,  $
\rho'(t,0) = P'(t,0) = \chi(t,0) = \Pi'(t,0) = \lambda'(t,0) = \phi'(t,0) = 0.$

\subsection{Cosmological decomposition}

To improve numerical accuracy and stability when solving the Misner–Sharp equations in an expanding universe, we factor out the homogeneous cosmological background from the full variables. This procedure is particularly effective since the Misner–Sharp formalism naturally captures the background spacetime evolution (see, e.g.,~\citep{Milligan:2025zbu}). With this in mind, we introduce a set of dimensionless variables:
\begin{subequations}\label{eqs:cosmoMD}
\begin{equation}
    R = aA\tilde{R},
\end{equation}
\begin{equation}
    \rho_{\rm pf} = \rho_{\rm T,b}\tilde{\rho}_{\rm pf}, \qquad \rho_\psi = \rho_{\rm T,b}\tilde{\rho}_\psi,
\end{equation}
\begin{equation}
    P_{\rm pf} = \rho_{\rm T,b}\tilde{P}_{\rm pf}, \qquad P_{\psi} = \rho_{\rm T,b}\tilde{P}_{\psi},
\end{equation}
\begin{equation}
    m_{\rm pf} = \frac{4\pi}{3}\rho_{\rm T,b}R^3\tilde{m}_{\rm pf}, \qquad m_{\psi} = \frac{4\pi}{3}\rho_{\rm T,b}R^3\tilde{m}_{\psi},
\end{equation}
\begin{equation}
    U = HR\tilde{U},
\end{equation}
\begin{equation}
    \psi = R_H\sqrt{\rho_{\rm T,b}}\,\tilde{\psi}, \qquad \chi = \sqrt{\rho_{\rm T,b}}\,\tilde{\chi}, \qquad \Pi = \sqrt{\rho_{\rm T,b}}\,\tilde{\Pi}.
\end{equation}
\end{subequations}
Here, $a$, $H$, and $\rho_{\rm T,b}$ denote the scale factor, the Hubble parameter, and the total background energy density, respectively. As previously discussed in Ref.~\citep{Milligan:2025zbu}, we assume that the background universe satisfies an effective barotropic equation of state $P_{\rm T,b} = w \rho_{\rm T,b}$, and define the initial Hubble radius as $R_H$. We also introduce a logarithmic time variable,
\begin{equation}
    \xi = \ln\left(\frac{t}{t_0}\right) = \frac{1}{\alpha} \ln a,
\end{equation}
where $\alpha = 2/3(1+w)$, and define the dimensionless radial coordinate $\bar{A} \equiv A / R_H$.

With this change of variables, the Misner–Sharp system transforms into the following set of evolution equations:
\begin{widetext}
\begin{subequations}\label{eq:final}
\begin{equation}\label{eq:SFevo}
    \partial_\xi\tilde{\psi} = \alpha e^{\phi+\xi}\tilde{\Pi} + \tilde{\psi},
\end{equation}
\begin{equation}\label{eq:chifinal}
    \partial_\xi\tilde{\chi} =  \alpha e^{\phi + \xi}(\tilde{\Pi}^{\prime} + \phi^{\prime}\tilde{\Pi}) + \tilde{\chi},
\end{equation}
\begin{equation}\label{eq:SFPIevo}
    \partial_\xi\tilde{\Pi}=\tilde{\Pi} +\alpha e^{\phi}\bigg\lbrace{ e^{\xi-\lambda}}\left[\left(\frac{2(\bar{A}\tilde{R})^{\prime}}{\bar{A}\tilde{R}} + \phi^{\prime} -\frac{\lambda^{\prime}}{2}\right)\tilde{\chi} + \tilde{\chi}^{\prime}\right] - \left[\left(2\tilde{U} + \frac{(\bar{A}\tilde{R}\tilde{U})^{\prime} + \frac{3}{2}e^{-\xi}\bar{A}\tilde{R}\tilde{\chi}\tilde{\Pi}}{(\bar{A}\tilde{R})^{\prime}}\right)\tilde{\Pi}+ e^{\xi}\tilde{V}(\tilde{\psi})_{,\tilde{\psi}}\right]\bigg\rbrace,
\end{equation}
\begin{equation}\label{eq:Rfinal}
    \partial_\xi\tilde{R} = \alpha\tilde{R}(\tilde{U}e^{\phi} - 1),
\end{equation}
\begin{equation}\label{eq:mfinal}
    \partial_\xi\tilde{\rho}_{\rm pf} = 2\tilde\rho_{\rm pf} -\alpha e^\phi\left[2\tilde U +\frac{(\bar{A}\tilde{R}\tilde{U})^{\prime} + \frac{3}{2}e^{-\xi}\bar{A}\tilde{R}\tilde{\chi}\tilde{\Pi}}{(\bar{A}\tilde{R})^{\prime}}\right](\rho_{\rm pf} + P_{\rm pf}),
\end{equation}
\begin{equation}\label{eq:Ufinal}
     \partial_\xi\tilde{U}= \tilde{U} - \alpha e^{\phi}\left[\frac{e^{2\xi - \lambda}(\bar{A}\tilde R)^{\prime}P^{\prime}_{\rm pf}}{(\bar{A}\tilde{R})(\tilde{\rho}_{\rm pf} + \tilde{P}_{\rm pf})} + \frac{3}{2}(\tilde{U}^2 + \tilde{P}_{\rm T}) +\frac{e^{2(1-\alpha)\xi}-\bar{\Gamma}^2}{2(\bar{A} \tilde{R})^2}\right],
\end{equation}
\begin{equation}
    \partial_\xi\lambda =\alpha e^\phi\left(\frac{2(\bar{A}\tilde{R}\tilde{U})^{\prime} + 3e^{-\xi}\bar{A}\tilde{R}\tilde{\chi}\tilde{\Pi}}{(\bar{A}\tilde{R})^{\prime}}\right),
\end{equation}
\begin{equation}\label{eq:lapseint}
    \phi^{\prime} = -\frac{\tilde{P}^{\prime}_{\rm pf}}{\tilde{\rho}_{\rm pf} + \tilde{P}_{\rm pf}},
\end{equation}
\begin{equation}\label{eq:rhoFinal}
    \tilde{\rho}_{\rm pf} = \tilde{m}_{\rm pf}  + \frac{\bar{A}\tilde{R}}{3(\bar{A}\tilde{R})^{\prime}}\tilde{m}^{\prime}_{\rm pf},
\end{equation}
\begin{equation}
    \tilde{\rho}_{\psi} = \tilde{m}_{\psi}  + \frac{\bar{A}\tilde{R}}{3(\bar{A}\tilde{R})^{\prime}}\tilde{m}^{\prime}_{\psi},
\end{equation}
\begin{equation}\label{eq:Gamma}
    \Bar{\Gamma}^2 = \frac{\Gamma^2}{a^2H^2R_H^2} = e^{2(1-\alpha)\xi} +\Bar{A}^2\tilde{R}^2(\tilde{U}^2 - \tilde{m}_{\rm T}) + \frac{3e^{-\xi}}{\Bar{A}\tilde{R}}\mathcal{I},
\end{equation}
\end{subequations}
\end{widetext}
where
\begin{equation}\nonumber
    \mathcal{I} = \int^{\bar A_0}_0\tilde{U}\bar A^3\tilde{R}^3\tilde{\chi}\tilde{\Pi} d\bar A.\nonumber   
\end{equation}
In the equations above, we repurposed primes to denote derivatives with respect to \( \bar{A} \).

To ensure regularity at the origin, we impose the following conditions at \( \bar{A} = 0 \):
\[
\tilde{\rho}_{\rm pf}^{\prime} = \tilde{R}^{\prime} = \tilde{m}_{\rm pf}^{\prime} = \tilde{U}^{\prime} = {\phi}^{\prime} = {\lambda}^{\prime} = \tilde{\chi} = \tilde{\Pi}^{\prime} = 0.
\]

\subsection{Initial conditions}


A systematic procedure for constructing initial conditions in the linear regime of a gradient expansion formalism—selecting only the growing mode and valid for super-horizon perturbations—was originally proposed by Shibata and Sasaki~\citep{Shibata_1999}. This method was later refined by Bloomfield \textit{et al.}~\citep{Bloomfield:2015} for a universe dominated by a perfect fluid, incorporating a double expansion: one in spatial gradients and the other in perturbation amplitudes. Following the methodology outlined in the latter, in Ref.~\citep{Milligan:2025zbu} we extended the framework to derive suitable initial conditions for both scalar field–dominated scenarios and the standard radiation-dominated case, valid to first order in both expansions.  In this work, we do not reproduce the full derivation again. Instead, we summarise the key steps that lead to the final initial data used in our simulations, along with a few additional modifications required to obtain the correct initial conditions in our specific setup. For a comprehensive treatment of the derivation, we refer the reader to Ref.~\citep{Milligan:2025zbu}.

\subsubsection{Initial conditions for a perfect fluid-dominated scenario}

The idea behind the derivation of our initial conditions is to linearize the equations of motion by expressing the variables as  
\begin{equation}\label{eq:linearisedX}
    \tilde{X} = \tilde{X}_{\rm b} + \epsilon\,\delta_X,
\end{equation}  
where \( \tilde{X} \) stands for the relevant dynamical quantities, such as \( \tilde{m}_{\rm pf} \), \( \tilde{U} \), \( \tilde{R} \), \( \tilde{\rho}_{\rm pf} \), etc.; \( \tilde{X}_{\rm b} \) denotes the corresponding background value\footnote{Note that most variables are unity at the background level.}, and \( \epsilon \) is a formal counting parameter that tracks perturbative order.  

In the absence of the scalar field—that is, in the perfect fluid dominated case—the Einstein equations can be manipulated to yield a second-order partial differential equation for the fluid mass perturbation:  
\begin{equation}\label{eq:dmPDE}
    \begin{split}
    \partial^2_\xi\delta_{m_{\rm pf}} - (3 - 5\alpha)\,\partial_\xi\delta_{m_{\rm pf}} + [(1-2\alpha)3w - 1]\,\alpha\,\delta_{m_{\rm pf}} \\
    = w\,\alpha^2\,e^{2(1-\alpha)\xi}\left(\frac{4\,\delta^\prime_{m_{\rm pf}}}{\Bar{A}} + \delta^{\prime\prime}_{m_{\rm pf}}\right).
    \end{split}
\end{equation}  
The right-hand side of this equation corresponds to spatial gradient terms and is suppressed in a gradient expansion. Neglecting this RHS contribution and retaining only the growing mode solution, we obtain the following expression for the initial perturbation:
\begin{subequations}
\begin{equation}\label{eq:dmsol}
    \delta_{m_{\rm pf}}(\Bar{A},\xi) = \delta_{{m0}_{\rm pf}}\,e^{2(1-\alpha)\xi}.
\end{equation}
Letting $\delta_{m0_{\rm pf}}=\delta_{m_{\rm pf}}(\Bar{A},\xi=0)=\delta_{m_{\rm pf}}(\Bar{A},t=t_0)$. With this expression for $\delta_{m_{\rm pf}}$, the remaining perturbative first order solutions are given by:
\begin{equation}
    \delta_R(\bar A,\xi) = -\frac{\alpha}{2}\left[\delta_{m0_{\rm pf}}+\frac{\omega}{1+3\omega}\bar A \delta_{m0_{\rm pf}}^\prime\right]e^{2(1-\alpha)\xi},        
\end{equation}
\begin{equation}
    \delta_{\rho_{\rm pf}}(\bar A,\xi) = \left[\delta_{_{m0_{\rm pf}}} + \frac{\Bar{A}}{3}\delta^\prime_{_{m0_{\rm pf}}}\right]e^{2(1-\alpha)\xi},
\end{equation}
\begin{equation}
    \delta_U (\bar A,\xi) =  -\frac{\alpha}{2}\delta_{m0_{\rm pf}}e^{2(1-\alpha)\xi}.
\end{equation}
\end{subequations}
As we noted above, these solutions are valid to first order in gradient and amplitude perturbations, and will serve as the basis for our numerical evolution. To fully specify the initial conditions of our system, it is also necessary to provide initial data for \( \lambda \). While this could be obtained by applying the same perturbative procedure used for the other variables, we instead choose to determine its initial value using the definition of \( \Gamma \) in Eq.~\eqref{eq:ugam}, followed by the relation \( \lambda = 2\ln(R'/\Gamma) \), where \( \Gamma \) is computed from the constraint equation \eqref{eq:Gam1} (or, equivalently, \eqref{eq:Gamma}).

\subsubsection{Initial conditions for a scalar field-dominated scenario}

Since our system is evolved in coordinates comoving with a perfect fluid, even in this scalar field-dominated scenario, we include a homogeneous, isotropic, and extremely diluted fluid component to regularize the origin and maintain the structure of the Misner--Sharp formalism. Perturbations are then attributed entirely to the scalar field and metric variables.

Unlike the perfect fluid-dominated scenario, in the scalar field-dominated case we were unable to find exact solutions to the Einstein equations that remain valid on superhorizon scales. Instead, we rely on approximate solutions. Following standard approaches~\citep{Polnarev_2012, Nakama_2014}, we relate perturbations to an initial curvature profile $K(A)$, which captures the growing mode at early times. This profile is directly linked to the local mass perturbation $\delta_{m_\psi}$ and allows us to express all other perturbed variables in terms of it. 

We assume that at the initial logarithmic time $\xi = 0$, the scalar field is purely kinetic, such that
\begin{equation}
    \tilde{\psi}_{0} = 0, \quad \tilde{\chi}_0 = 0, \quad \tilde{\Pi}_{\rm b0}^2 / 2 = 1-\rho_{\rm b0_{pf} },
\end{equation}
where $\rho_{\rm b0_{pf} }$ is the initial contribution of the extremely diluted perfect fluid. Under these assumptions, the energy density perturbation takes a simplified form, from which we derive the following initial conditions, valid to first order in gradient and amplitude perturbations:
\begin{align}
    \delta_{\Pi 0} &= \frac{1}{\sqrt{2}}\left(\delta_{m0_\psi} + \frac{\bar{A}}{3} \delta^\prime_{m0_\psi} \right), \\
    \delta_{U0} &= -\frac{n+1}{6n}\left(4 \delta_{m0_\psi} + \bar{A} \delta^\prime_{m0_\psi} \right), \\
    \delta_{R0} &= -\frac{(n+1)^2}{12n(2n - 1)}\left(4 \delta_{m0_\psi} + \bar{A} \delta^\prime_{m0_\psi} \right).
\end{align}
In the above expression, $n$ is related to the exponent of the scalar field potential, such that $V(\psi)\propto \psi^{2n}$.

Since $\delta_{m0_\psi}$ is directly related to the curvature perturbation $\tilde{K}$, we use it as the fundamental initial profile in our simulations. To determine the initial condition for $\lambda$, we follow the same procedure as in the radiation-dominated scenario.

\section{Trapping horizon and postcollapse dynamics}\label{Sec:III}

\subsection{Trapping horizons and condition for primordial black hole formation}

To determine when a black hole forms within our simulations, we rely on the identification of trapped surfaces—regions in which the expansion of outgoing null rays becomes negative. These surfaces indicate that light cannot escape in the outward direction, signaling the presence of a local trapping horizon and, consequently, black hole formation~\citep{Helou_2017}.

This identification is made by evaluating the expansion of radial null geodesics, defined as
\begin{equation}
    \Theta^{\pm} = h^{\mu\nu} \nabla_{\mu} k^{\pm}_{\nu},
\end{equation}
where $k^{\pm}_{\nu}$ denote the future-directed outgoing and ingoing null vectors. The induced metric on a spherical surface is given by
\begin{equation}
    h^{\mu\nu} = g^{\mu\nu} + n^\mu n^\nu - s^\mu s^\nu,
\end{equation}
where $n^\mu$ is a unit timelike vector orthogonal to constant-time hypersurfaces, and $s^\mu$ is a unit spacelike vector orthogonal to surfaces of constant radial coordinate $A$. These vectors allow us to construct the outgoing and ingoing null vectors, which, in our coordinates, takes the explicit form
\begin{equation}
    k^{\mu(\pm)} = \frac{1}{\sqrt{2}}(e^{-\phi}, \pm e^{-\lambda/2}, 0, 0).
\end{equation}

In a spherically symmetric spacetime, the expansion reduces to
\begin{equation}
    \Theta^{\pm} = \frac{1}{4\pi R^2} k^{\mu(\pm)} \nabla_{\mu}(4\pi R^2),
\end{equation}
which simplifies to
\begin{equation}
    \Theta^{\pm} = \frac{\sqrt{2}}{R}(U \pm \Gamma).
\end{equation}
Expressed in terms of dimensionless variables, this becomes
\begin{equation}
    \Theta^{\pm} = \sqrt{2}H\left(\tilde{U} \pm \frac{\bar{\Gamma}}{\bar{A} \tilde{R}}\right).
\end{equation}

We identify black hole formation by locating an apparent horizon, which occurs when $\Theta^+ = 0$ and $\Theta^- < 0$. A cosmological (anti-trapped) horizon corresponds to $\Theta^- = 0$ with $\Theta^+ > 0$, while a bifurcation surface—representing a time-symmetric trapping horizon—occurs when $\Theta^+ = \Theta^- = 0$.

\subsection{The compaction function}

A particularly useful tool for studying the dynamics of cosmological perturbations is the compaction function, $ \mathcal{C} $, which quantifies the excess energy carried by the radiation fluid (or scalar field) within a given region. In terms of our formalism, it can be conveniently computed using the following expression:
\begin{equation}
    \mathcal{C}(\bar{A}) \equiv \tilde{A}^2 \tilde{R}^2(\tilde{A}) \left[ \tilde{m}_{\rm T}(\tilde{A}) - \tilde{m}_{\rm T,b}(\tilde{A}) \right] e^{2(\alpha - 1)\xi},
\end{equation}
where 
$ \tilde{m}_{\rm T,b} = \frac{4\pi}{3} \tilde{\rho}_{\rm T,b} \tilde{R}^3 $, and 
$ \tilde m_{\rm T} = \tilde m_{\rm pf} $ or $ \tilde m_{\rm T} = \tilde m_{\psi} $ in the radiation- or scalar field-dominated scenarios, respectively.

The qualitative behavior of the maximum of the compaction function provides important insight into the fate of the perturbation. If $\mathcal{C}_{\rm max}$ grows and exceeds a certain threshold (typically close to unity), this is indicative of gravitational collapse and the potential formation of a PBH. Conversely, if $\mathcal{C}_{\rm max}$ decreases over time, this means the perturbation is not dense enough to collapse and will eventually disperse.

While throughout this work we continue to define PBH formation by the appearance of a {future trapped surface}, specifically via the detection of an {apparent horizon}, we also present the time evolution of the maximum of the compaction function as a complementary diagnostic. Its behavior provides a physically intuitive and computationally inexpensive indicator of collapse or dissipation, offering additional insight even when an apparent horizon has not yet fully formed.

\subsection{Primordial black hole mass and causality condition}\label{Sec:PBHdynamics}

To evaluate the mass of the PBH, we employ the Misner--Sharp mass, a quasi-local quantity that measures the total energy enclosed within a spherical region of areal radius $R$. This Misner--Sharp mass is defined geometrically as
\begin{equation}\label{eq:MSmass}
    M_{\rm MS} = \frac{R}{2} \left(1 - \nabla^a R \nabla_a R \right).
\end{equation}
Using our formalism and the dimensionless variables introduced earlier, the above equation becomes
\begin{equation}
    \label{eq:MS_mass}
    M_{\rm MS} = \frac{e^{\alpha \xi} \bar A \tilde R R_H}{2} \left[1 + e^{2(\alpha - 1)\xi} \left((\bar A \tilde R \tilde U)^2 - \bar \Gamma^2 \right) \right].
\end{equation}

To further probe the dynamics and geometry of PBH formation, we also calculate the velocity of the apparent horizon. Within our formalism, and using the rescaled variables, this velocity is given by (see Appendix~\ref{App:velocity} for details):
\begin{equation}
    \label{eq:VAH_tilde}
    v_{\rm AH} = \frac{\alpha + 1}{\alpha - 1},
\end{equation}
with the parameter \( \alpha \) defined as
\begin{equation}\label{eq:alpha}
    \alpha = \frac{(\bar A \tilde R)^2 (\tilde \rho_T + \tilde P_T + 2\tilde \chi \tilde \pi\, e^{-\lambda/2})}{\frac{2}{3}e^{2(1 - \alpha)\xi} - (\bar A \tilde R)^2 (\tilde \rho_T - \tilde P_T)}.
\end{equation}

The value of \( v_{\rm AH} \) determines the causal nature of the apparent horizon \citep{Helou:2016xyu}:
\begin{itemize}
    \item If \( v_{\rm AH} > 1 \), the horizon is spacelike, expanding faster than the local speed of light. This indicates strong infall and rapid black hole growth.
    
    \item If \( v_{\rm AH} = 1 \), the horizon is null and propagates along the outgoing light cone. This corresponds to a nearly stationary black hole with minimal ongoing accretion.
    
    \item If \( v_{\rm AH} < 1 \), the horizon would be timelike, which is generally unphysical in realistic collapse scenarios, as it would suggest the possibility of information escaping from within the horizon.
\end{itemize}

Once the apparent horizon has formed, continued numerical evolution typically becomes infeasible due to the emergence of singularities in the interior region. To address this, we adopt an excision technique \citep{PhysRevLett.69.1845, PhysRevD.51.5562} (see also Refs.~\citep{Escriva:2019nsa,Escriva:2025eqc} for recent applications in the Misner–Sharp formalism) that removes a portion of the computational domain enclosed by the apparent horizon. This approach is justified by the causal structure of spacetime: the idea is that since information cannot propagate from within the horizon to the exterior, excluding the interior should not affect the evolution of the outer region.

Our excision procedure follows the methodology outlined below. We first define the parameter $D \bar A = n\,d\bar A$, where $d\bar A$ is the spatial resolution of the simulation and $n \geq 1$ is an integer controlling the distance between the apparent horizon and the new inner boundary of the computational domain. The system is then evolved until the distance between the apparent horizon (which evolves dynamically to larger values of $\bar A$) and the current inner boundary reaches $\Delta \bar A = m\,d\bar A$, with $m > n$. In our simulations, we fixed $n=1$ and $m=2$. At that point, the domain is excised again, updating the inner boundary to lie once more at a distance of $n$ grid points from the horizon. This cycle is repeated to track the post-collapse evolution of the system while maintaining numerical stability and physical accuracy in the exterior region.

\section{Numerical methodology and consistency check}\label{Sec:IV}

\subsection{Implementation of the numerical scheme}

For our numerical simulations, we used a modified version of the numerical code we previously introduced in Ref.~\citep{Milligan:2025zbu}. The method used to solve system~\eqref{eq:final} numerically is described below. Time evolution is carried out using the method of lines: the dynamical variables 
\( (\tilde\psi, \tilde\chi, \tilde\Pi, \tilde R, \tilde \rho_{\rm pf}, \tilde U, \lambda) \) 
are evolved using a fourth-order Runge--Kutta integrator. The evolution proceeds in discrete steps along the logarithmic time coordinate \( \xi \), such that \( \xi \rightarrow \xi + \Delta \xi \). At each step, both the evolution and constraint equations are updated.

Spatial derivatives are computed using fourth-order accurate finite difference methods, \( \mathcal{O}(\Delta \Bar{A}^4) \). Before any excision is applied, we exploit the symmetry of the system to impose reflective boundary conditions at the origin, ensuring that first derivatives vanish there. After excision, numerical derivatives at the excision boundary are computed using a seven-point outward finite difference stencil. At the outer boundary, we explicitly impose Neumann boundary conditions at all times:
\begin{equation}
    \frac{\partial f}{\partial \Bar{A}} = 0,
\end{equation}
for any evolving variable $f$.

Special care must be taken with Eq.~\eqref{eq:Gamma}, since one might be tempted to use this relation to compute $\Gamma$ and then substitute it into the other differential equations (such as Eq.~\eqref{eq:Ufinal}). However, our approach was instead to calculate $\Gamma$ directly from its definition in Eq.~\eqref{eq:ugam}, while using Eq.~\eqref{eq:Gamma} solely as a consistency check for numerical accuracy (see next subsection).

High-order finite difference methods can be susceptible to spurious numerical oscillations, especially due to steep gradients or interpolation noise. To mitigate such effects, we incorporate a third-order Kreiss–Oliger dissipation term~\citep{Kreiss}, which suppresses high-frequency modes. The evolution equations for all variables \( f \) are modified as follows:
\begin{equation}
    \begin{split}
    \partial_\xi f \rightarrow& \partial_\xi f \\&+ \frac{\tilde \sigma}{64\Delta\Bar{A}}(f_{i+3} - 6f_{i+2} + 15f_{i+1} \\
    & - 20f_i + 15f_{i-1} - 6f_{i-2} + f_{i-3}),
    \end{split}
\end{equation}
where \( i \pm j \) refers to grid point offsets and \( \tilde\sigma \sim \mathcal{O}(10^{-2}) \) controls the strength of dissipation.

Throughout all simulations, we set the initial scale factor \( a(t_0) = 1 \) and logarithmic time \( \xi_0 = 0 \), with time related to the scale factor via \( t = t_0 e^\xi \). The effective background fluid equation of state is taken to be \( w = 1/3 \), which is appropriate for both a radiation-dominated universe and a scalar field dominated by a quartic potential.\footnote{It is important to note that this equation of state is only an effective one. The full dynamics—such as the presence of oscillations in the Misner–Sharp variables, which are expected in scenarios dominated by scalar fields—will be reflected in our rescaled (primed) variables.} This choice determines the Hubble scale and horizon radius at the initial time: \( H_0 = 1/(2t_0) \) and \( R_H = 2t_0 \). We also adopted the parameter value $\hat \lambda_4 \equiv R_{H}^2 \lambda_4 = 10$ for the quartic self-interaction strength of the scalar field in all our simulations. 

As the initial profile for the overdensity, we adopted a simple Gaussian shape for $\delta_{m_{\rm T}}$ (where, as stated before, $\delta_{m_{\rm T}} = \delta_{\rm m0_{pf}}$ in the radiation collapse scenario and $\delta_{m_{\rm T}} = \delta_{\rm m0_\psi}$ in the scalar field collapse case), given by:
\begin{equation}
    \delta_{m_{\rm T}} = p\, e^{-A^2 / 2\Sigma^2},
\end{equation}
where, for simplicity, we fixed $\Sigma = 5R_{\rm H}$ in all simulations. This left $p$ as the only free parameter in our setup.

Special care has been taken when determining the initial condition for $\lambda$ in the scalar field-dominated scenario. This is because calculating $\lambda$ requires knowing the value of $\Gamma$, which in turn depends on the mass associated to the scalar field contribution. The mass is obtained by integrating the energy density of the scalar field, but this density itself depends on $\Gamma$, making the problem inherently circular. We resolve this interdependence through an iterative procedure at the initial time step, which allows us to converge to a consistent value of $\Gamma$, and consequently, to the correct initial value of $\lambda$.

To identify PBH formation, we monitor the computational grid for the conditions specified in Sec.~\ref{Sec:PBHdynamics}, particularly the criterion $\theta^+ = 0$. Once this condition is satisfied, we initiate the excision of the domain interior to the apparent horizon and begin tracking the evolution of the horizon. Specifically, we compute the Misner--Sharp mass and the horizon's velocity using Equations \eqref{eq:MS_mass} and ~\eqref{eq:VAH_tilde}, respectively. To assess whether the PBH has reached its final mass, we examine the apparent horizon velocity, $v_{\rm AH}$. If $v_{\rm AH} \approx 1$ within a tolerance $\delta$, we consider the PBH mass to have stabilized.

\subsection{The Hamiltonian constraint and the criterion for convergence}

To assess the numerical accuracy of our results, we monitor the consistency of the Hamiltonian constraint given by Eq.~\eqref{eq:Gamma}. If the Einstein equations are correctly solved, this quantity should remain small throughout the evolution.\footnote{In fact, in an ideal case, this quantity would be exactly zero.} However, once the excision method is initiated, Eq.~\eqref{eq:Gamma} can no longer be evaluated due to the removal of part of the computational domain, which makes it impossible to compute the corresponding integrals.

To continue tracking the accuracy of the simulation beyond this point, we instead adopt an alternative—but equivalent—formulation of Eq.~\eqref{eq:Gamma}, derived in Appendix~\ref{App:HC}. For completeness, we reproduce here the explicit expression for the Hamiltonian constraint used in our computations:
\begin{eqnarray}\label{eq:HC}
   {\mathscr{H}} &=& \bar A^2 \tilde R^2 \tilde U\left[{(\bar A\tilde{R}\tilde{U})^{\prime} + \frac{3}{2}e^{-\xi}\Bar{A}\tilde{R}\tilde{\chi}\tilde{\Pi}}\right] \nonumber \\
    && - {(\bar A\tilde R)^{\prime}}\left[\frac{\bar A\tilde R\bar \Gamma'}{e^{\lambda/2-\xi}}+\frac{3}{2}\bar A^2\tilde R^2\tilde \rho_{\rm T} \right. \\
    &&\left.+ \frac{\bar \Gamma^2 -(\bar A\tilde R\tilde U)^2 - e^{2(1-\alpha)\xi}}{2}\right].\nonumber 
\end{eqnarray}
To quantify the accuracy, we compute a weighted measure Hamiltonian constraint, defined as:
\begin{equation}
    \mathcal{H} = \frac{|\mathscr{H}|}{\sum_i |T_i|},
\end{equation}
where \( T_i \) denotes each of the individual terms contributing to \( \mathscr{H} \), i.e., \( \mathscr{H} = \sum_i T_i \).

\section{Critical behavior in radiation and scalar field dominated scenarios}\label{Sec:V}

\subsection{The radiation-dominated case}

As a consistency check, and before analyzing the critical behavior in the scalar field-dominated case, we begin this section by reproducing the well-known results for critical collapse in a radiation-dominated universe. This scenario has been extensively studied in the literature, with critical behavior confirmed over several orders of magnitude near the threshold for PBH formation \citep{Musco:2008hv}.
Our purpose here is not to replicate the detailed analyses already presented in previous works, but rather to verify that our numerical code correctly captures this critical behavior. This serves both as a validation of our implementation and as evidence that our code reliably reproduces established numerical results.

\begin{figure}
    \centering
    \includegraphics[width=3.5in]{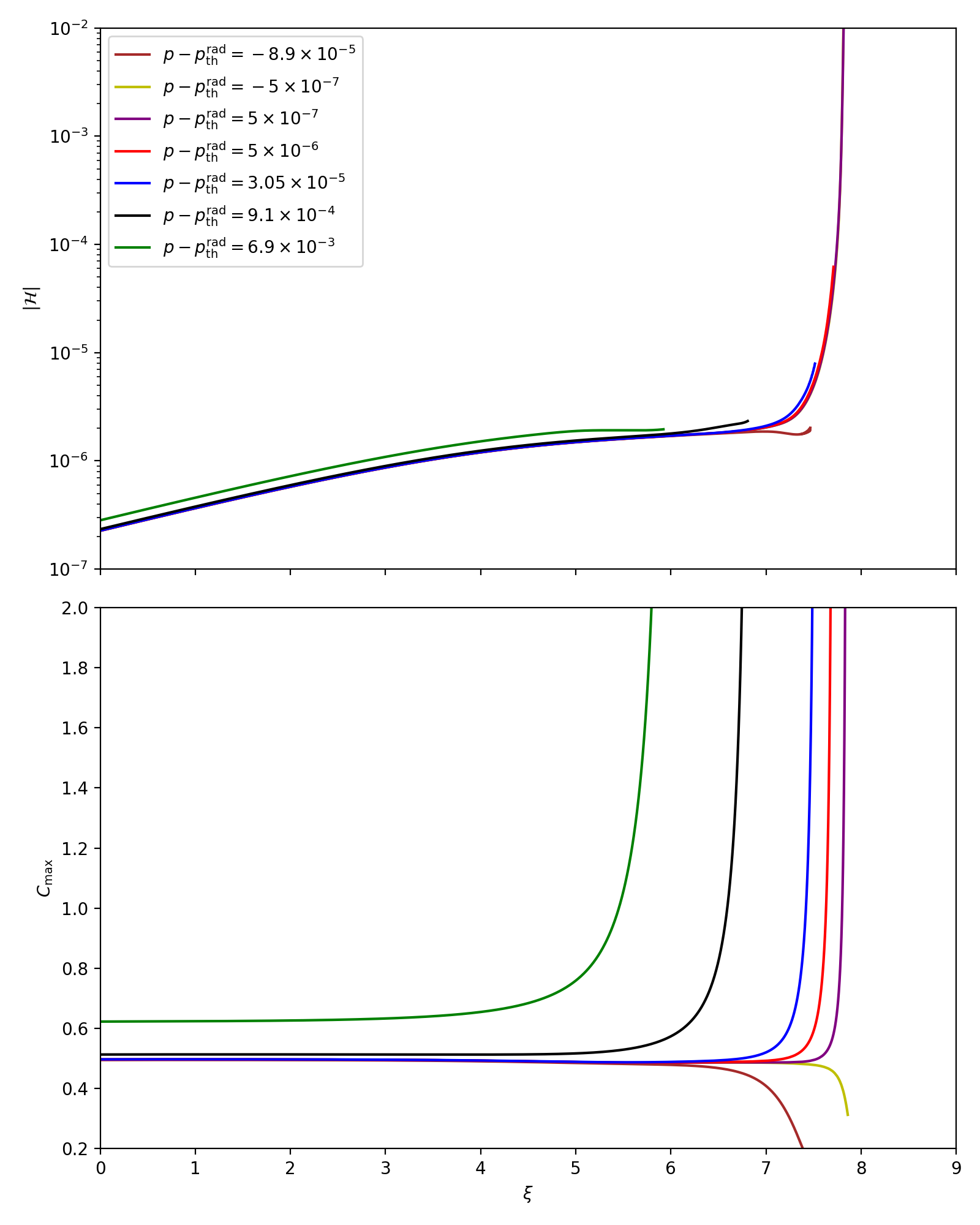}
    \caption{\footnotesize{Upper panel: Evolution of the Hamiltonian constraint $L^2$-norm for different initial conditions in a radiation-dominated universe. The results shown correspond to the phase before any excision is applied. Lower panel: Evolution of the maximum of the compaction function for the same set of initial conditions. Resolution: $d\bar{A} = 0.05$.}}
    \label{fig:HCrad}
\end{figure}

As a first step to achieve a reliable determination of critical behavior, it is essential to accurately estimate the threshold value for PBH formation. We define this threshold as
\begin{equation}
    p_{\rm th} = \frac{p^{\rm min}_{\rm bh} + p^{\rm max}_{\rm no-bh}}{2},
\end{equation}
where, in our simulations, \( p^{\rm min}_{\rm bh} \) is the smallest value of \( p \) for which a black hole forms, and \( p^{\rm max}_{\rm no-bh} \) is the largest value for which the perturbation disperses. Using this definition, we find the threashold value
\begin{equation}
p_{\rm th}^{\rm rad} = 0.0270895.
\end{equation}
Note that this value differs from that reported in Ref.~\citep{Milligan:2025zbu} because we now adopt an initial profile with $\Sigma = 5R_{\rm H}$, instead of the $\Sigma = 2R_{\rm H}$ used in our previous work. Nevertheless, when the threshold is obtained in terms of the maximum of the compaction function evaluated at horizon crossing (see lower panels of Figs.~\ref{fig:HCrad}), we find the threshold value $C_{\rm max, th} \simeq 0.5$, consistent with the typical result reported in the literature for an initial Gaussian profile.

Figure~\ref{fig:HCrad} shows the time evolution of the Hamiltonian constraint, measured using the $L^2$-norm (top panel), together with the maximum of the compaction function (bottom panel) for various initial values of $p$. These plots demonstrate that the compaction function provides a clear distinction between collapsing and dispersing configurations, and that the constraint remains well controlled during the early stages of evolution, showing significant growth at the final time steps only for initial conditions very close to the critical collapse threshold.

\begin{figure}
    \centering    \includegraphics[width=3.5in]{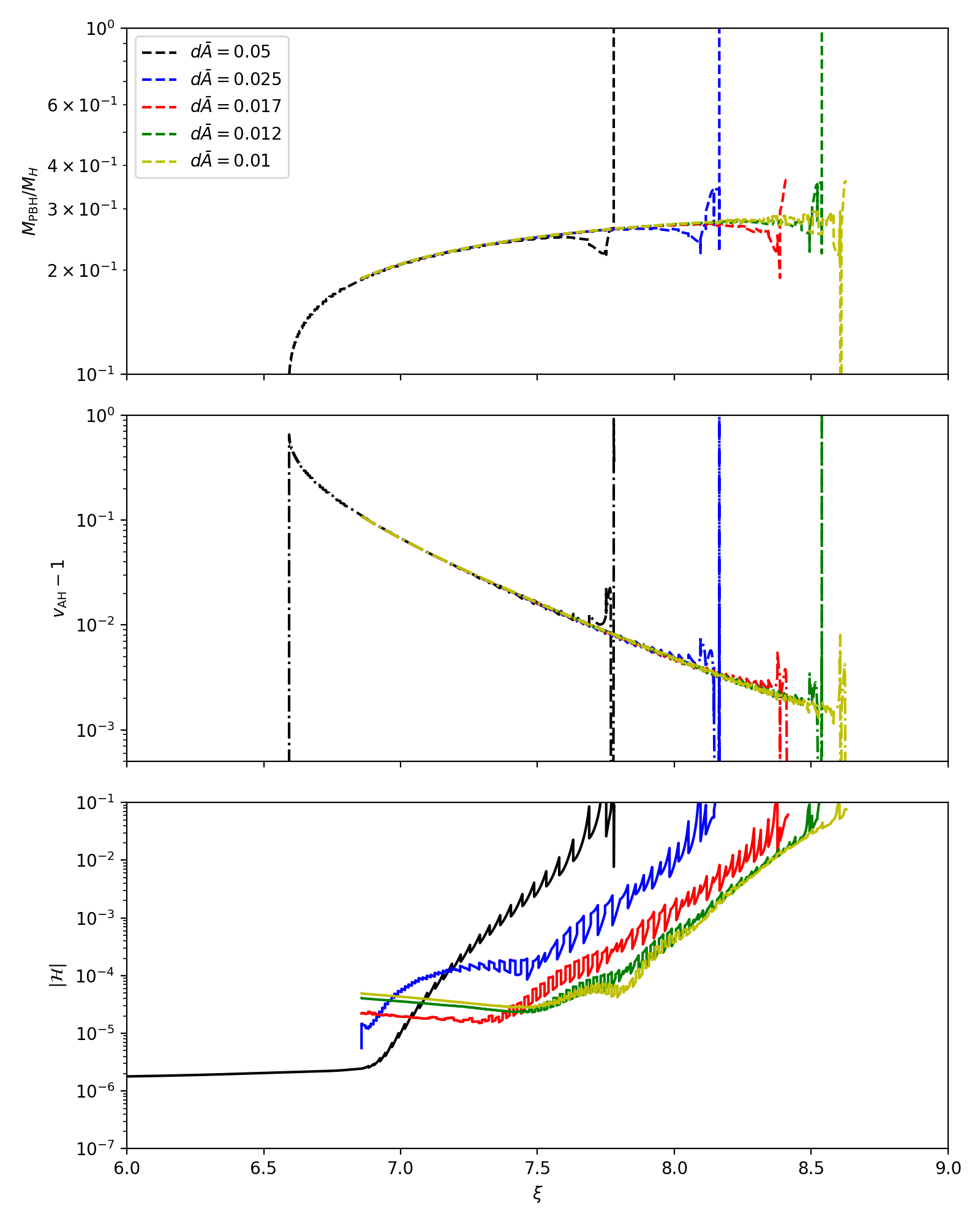}
    \caption{\footnotesize{Upper panel: Mass of the PBH divided by the cosmological horizon mass evaluated at horizon crossing, shown as a function of time for different resolutions.
Middle panel: Velocity of the apparent horizon as a function of time for the same set of resolutions. Bottom panel: Hamiltonian constraint evaluation. In this plot, we consider the specific case with $p - p_{\rm th} = 9.1 \times 10^{-4}$. }}
    \label{fig:Mandc_rad}
\end{figure}

An essential step in characterising the critical nature of the collapse is to accurately determine the mass of the resulting PBHs for different values of the parameter \( p \). To this end, in Fig.~\ref{fig:Mandc_rad} we focus on analyzing how the mass, the velocity of the apparent horizon, and the Hamiltonian constraint vary with resolution, in order to identify the level of refinement required to obtain reliable results. In this figure, we initially evolve the system using a resolution of \( d\bar{A} = 0.05 \) up to shortly after the formation of an apparent horizon, and subsequently extrapolate the results to higher resolutions to start with the excision implementation.

As can be seen from the figure, as the resolution of our numerical simulations is increased, the PBH mass approaches an asymptotic value, and simultaneously, the velocity of the apparent horizon tends to be null. We can observe that deviations from the expected asymptotic behavior at each resolution appear when the Hamiltonian constraint exceeds \( 10^{-2} \). Consequently, we adopt \( |\mathcal{H}| < 10^{-2} \) as a conservative threshold for trusting the simulation results.

It is also evident from the figure that when the velocity of the apparent horizon satisfies \( v_c - 1 < 3\times10^{-3} \), determining the PBH mass becomes increasingly challenging. This is due to the fact that the mass is already very close to its asymptotic value, and our numerical code struggles to resolve the final stages of evolution in this regime. For this reason, throughout our analysis we adopt the criteria \( v_c - 1 < 3\times10^{-3} \), $|\mathcal{H}|<10^{-2}$ and require that the mass exhibits well-behaved asymptotic convergence in order to report a final PBH mass.

\begin{figure}
    \centering
    \includegraphics[width=3.7in]{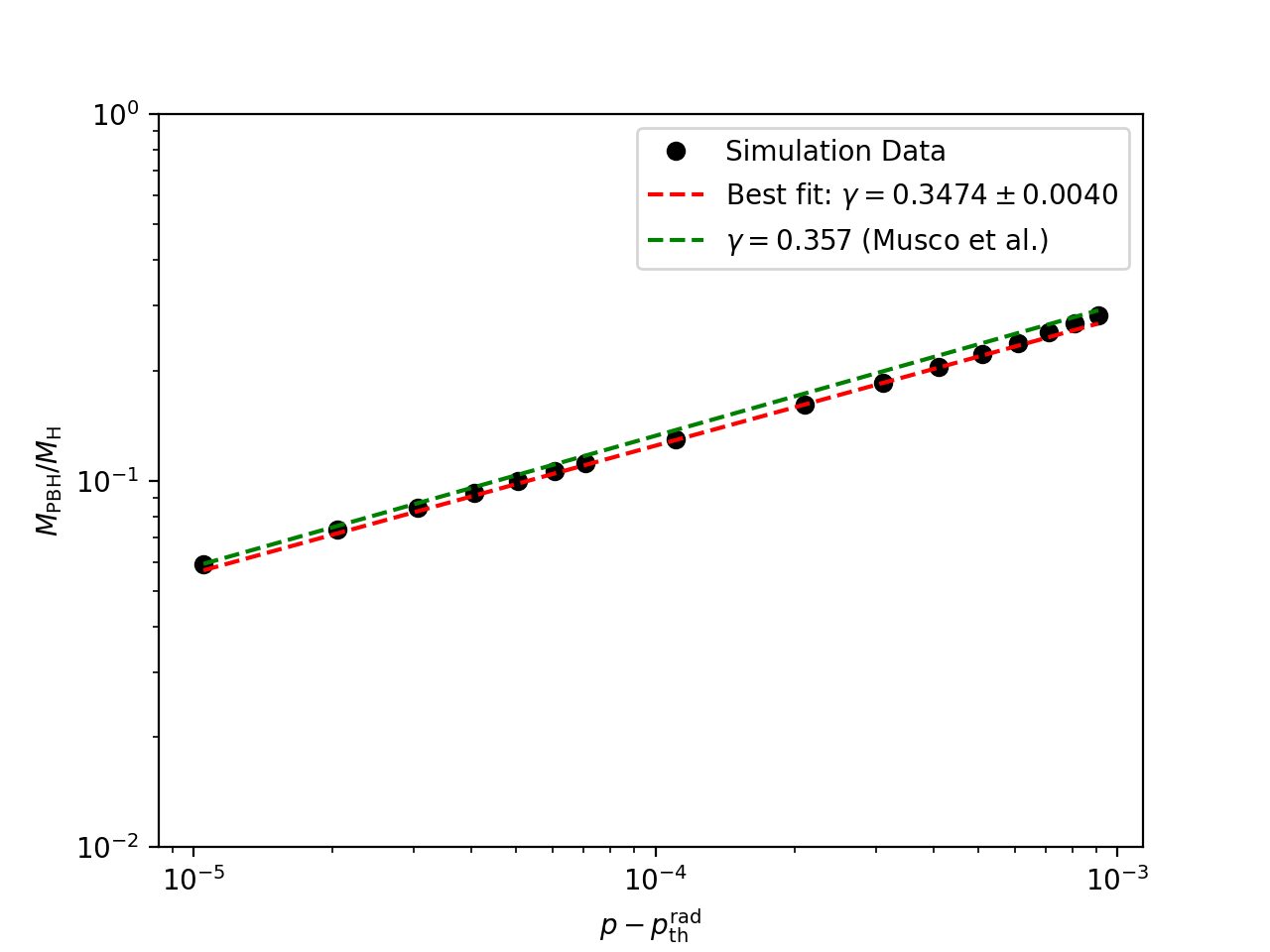}
    \caption{\footnotesize{Critical behavior of the PBH mass as a function of the distance to the critical threshold, $p - p_{\rm th}$. The data follow the expected power-law scaling $M_{\rm PBH} \propto (p - p_{\rm th})^\gamma$, with a best-fit critical exponent $\gamma = 0.3474 \pm 0.004$. For comparison, the value $\gamma = 0.357$ reported by Musco et al.~\citep{Musco:2008hv} for a radiation-dominated collapse is shown as a dashed green line. The small discrepancy is likely due to differences in numerical methods, and excision strategies, but remains within a few percent, confirming the robustness of the critical scaling behavior.}}

    \label{fig:crit_radiation}
\end{figure}

In Fig.~\ref{fig:crit_radiation}, we present the critical behavior obtained for different initial conditions, together with the corresponding fit of the scaling law. From our analysis, we find a best-fit critical exponent of \( \gamma = 0.3474 \pm 0.004 \), which is slightly lower than the value \( \gamma = 0.357 \) reported by Musco et al.~\citep{Musco:2008hv}.  
This small discrepancy can be attributed to differences in the numerical methods and resolution criteria adopted in each study. In particular, our analysis uses an excision strategy, which may influence the dynamics near criticality and the precise extraction of asymptotic quantities. Despite these differences, our result agrees each other, confirming the robustness of the critical behavior observed in radiation-dominated collapse.

\subsection{The scalar field-dominated case}

We now apply the same analysis to the scalar field–dominated scenario. Despite the distinct physical nature of the scalar field, our results confirm that the overall critical behavior—including the threshold for PBH formation and the critical exponent—closely resembles the radiation-dominated case, a conclusion already suggested in our previous work~\citep{Milligan:2025zbu}. 

\begin{figure}
    \centering
     \includegraphics[width = 3.5in]{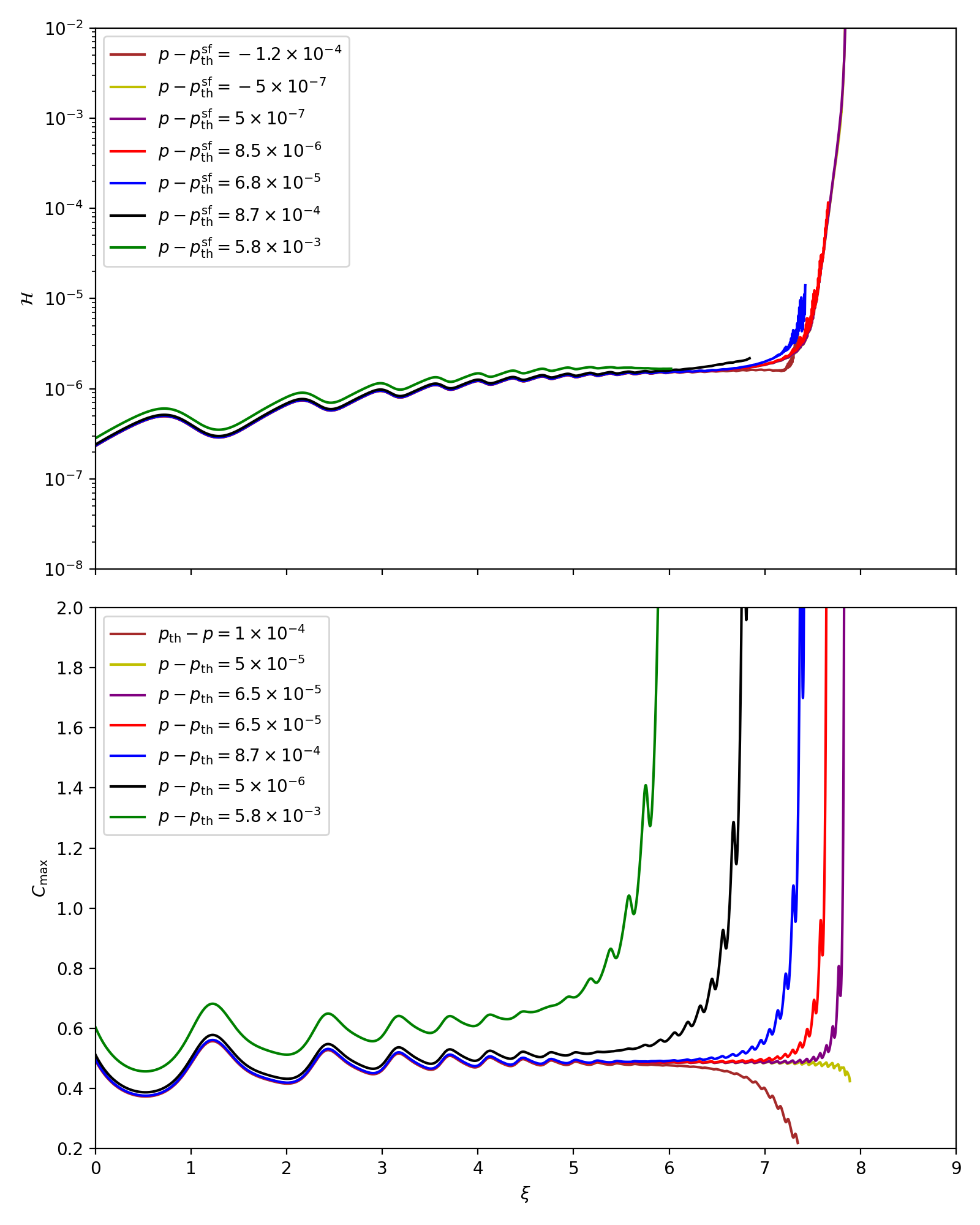}
    \caption{\footnotesize{Upper panel: $L^2$-norm of the Hamiltonian constraint evolution for different initial conditions in the scalar field-dominated case, before excision.  
    Lower panel: Evolution of the maximum of the compaction function for the same set of initial conditions. Resolution: $d\bar{A} = 0.05$.}}
    \label{fig:HCsf}
\end{figure}

Figure~\ref{fig:HCsf} shows the evolution of the Hamiltonian constraint and the maximum of the compaction function for different initial conditions. As before, the compaction function clearly distinguishes between collapsing and dispersing configurations, while the Hamiltonian constraint remains well controlled, showing significant growth only near criticality. We note, however, that in the scalar field case the maximum of the compaction function exhibits oscillations throughout the evolution, which makes it somewhat challenging to precisely determine the collapse threshold based on this quantity alone. Nevertheless, for our purposes—focused on exploring the critical nature of the collapse—we report the threshold in terms of the initial amplitude parameter \( p \), for which we find
\begin{equation}
    p_{\rm th}^{\rm sf} = 0.0271205.
\end{equation}

\begin{figure}
    \centering
    \includegraphics[width=3.5in]{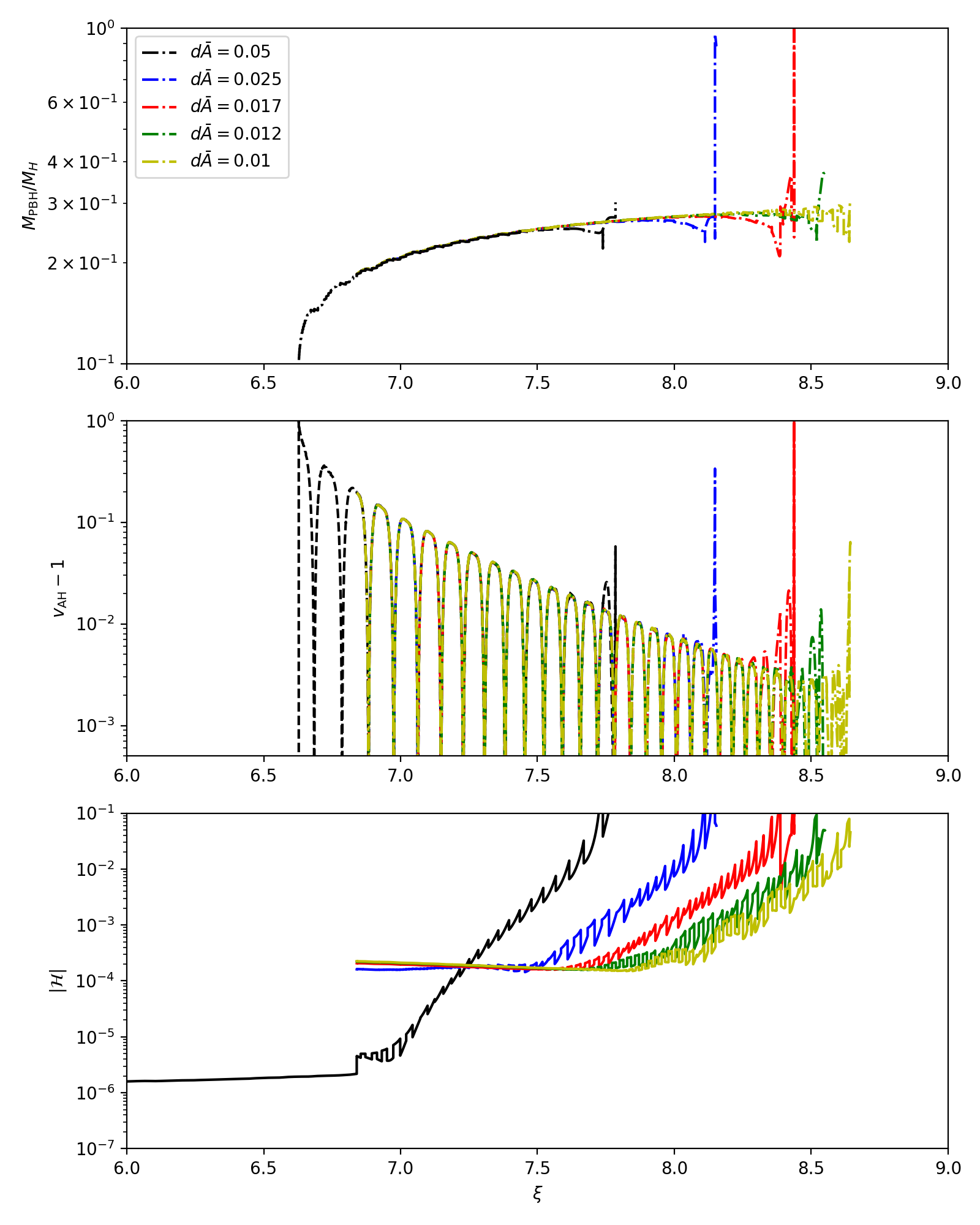}
    \caption{\footnotesize{Upper panel: PBH mass normalized by the cosmological horizon mass at horizon crossing, as a function of time for different resolutions.  
    Middle panel: Apparent horizon velocity as a function of time.  
    Bottom panel: Hamiltonian constraint evaluation. The case shown corresponds to $p - p_{\rm th} = 8.7 \times 10^{-4}$.}}
    \label{fig:Mandc_sf}
\end{figure}

We then examine the dependence of the PBH mass, the apparent horizon velocity, and the Hamiltonian constraint on numerical resolution in Fig.~\ref{fig:Mandc_sf}, aiming to assess whether the scalar field dynamics introduce any notable differences compared to the radiation case. The observed behavior is analogous: the PBH mass converges toward an asymptotic value, with deviations appearing when the Hamiltonian constraint exceeds \( 10^{-2} \). The apparent horizon velocity once again serves as a reliable indicator for identifying the asymptotic regime. Although this velocity shows oscillations, its overall amplitude decreases over time, signaling that the formed PBH is approaching a stationary state.

\begin{figure}
    \centering
    \includegraphics[width=3.7in]{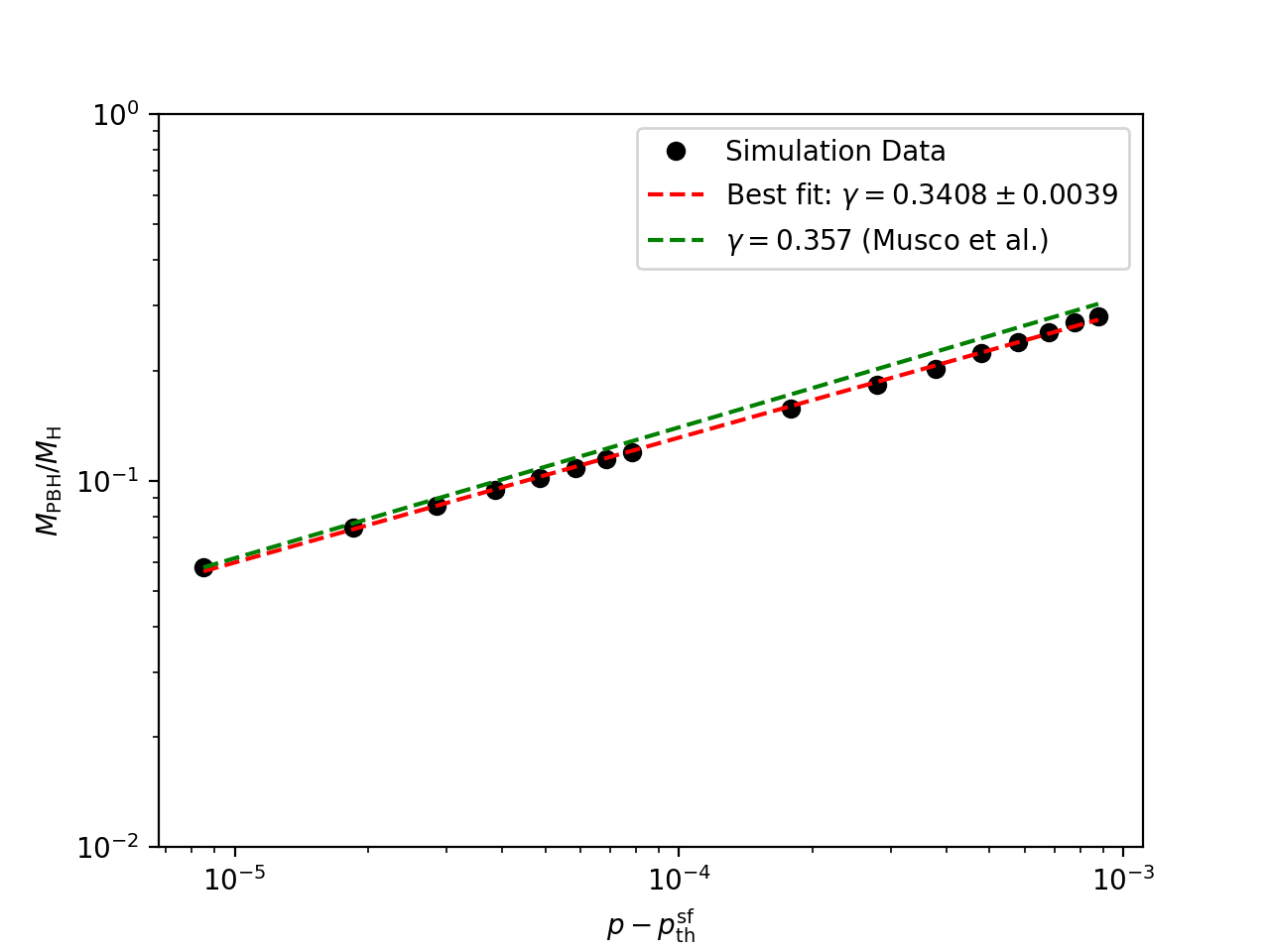}
    \caption{\footnotesize{Critical behavior of the PBH mass as a function of $p - p_{\rm th}$ in the scalar field-dominated case. The best-fit scaling yields $\gamma = 0.3401 \pm 0.0071$. The value $\gamma = 0.357$ reported by Musco~\citep{Musco:2008hv} for a radiation-dominated collapse is shown for comparison.}}
    \label{fig:crit_sf}
\end{figure}

Finally, Fig.~\ref{fig:crit_sf} presents the critical behavior of the PBH mass for the scalar field case. The data exhibit the expected power-law scaling, yielding a best-fit critical exponent of \( \gamma = 0.3401 \pm 0.0071 \). This value is close --- though different within \( 2\sigma \) --- to the result obtained for the radiation-dominated case. Our findings confirm that both scenarios display a remarkably similar type II critical behavior, indicating that, within the precision of our simulations, the critical exponent shows only mild dependence on whether the collapse is driven by a radiation fluid or a scalar field. This result further supports the near universality of the critical exponent across different matter models.

\section{Conclusions}\label{Sec:VII}

In this work, we performed a detailed numerical study of primordial black hole (PBH) formation in two distinct cosmological scenarios: a radiation-dominated universe and a universe dominated by a quartic self-interacting scalar field. Our goal has been to test whether the well-known critical behavior observed in radiation collapse extends to the scalar field case, particularly near the threshold of PBH formation.

We validated our numerical approach by reproducing known results for the radiation-dominated case, obtaining results consistent with those reported in the literature. We then applied the same analysis to the scalar field case and found that, despite the intrinsic differences in the underlying matter content, the critical behavior of PBH formation remains remarkably similar. Both the critical exponent and the threshold for collapse are in close agreement with those of the radiation scenario, although we observe a mild discrepancy in the critical exponent at the level of about \( 2\sigma \). A further distinction is the presence of oscillations in the physical quantities characterizing the system in the scalar field case.

This difference could stem from physical effects inherent to the scalar field dynamics ((such as the oscillatory nature of the field) or from limitations in numerical resolution and accuracy, particularly near the critical threshold where simulations are most sensitive. 

Overall, this study provides direct evidence that a quartic scalar field behaves effectively like a radiation fluid, even in the highly nonlinear regime of gravitational collapse. By contrast, in our previous analysis of the quadratic case \citep{Milligan:2025zbu} we found that this correspondence breaks down, with gradient effects producing significant departures from dust-like behavior. The critical behavior observed here therefore appears largely insensitive to the specific matter model, at least within the quartic scalar and radiation-dominated cases, reinforcing the near universality of type~II critical phenomena in PBH formation.

These findings lend support to the common practice of modeling certain scalar field scenarios with perfect fluids, provided the potential is chosen appropriately. At the same time, they highlight the need to test this correspondence directly in the nonlinear regime, since even subtle deviations can have important consequences for cosmology and the early universe.

\section{Acknowledgments}
We would like to thank Prof. Tomohiro Harada for many illuminating discussions. LEP is supported by a Royal Society funded post-doctoral position. EM is supported by an STFC studentship. DJM was supported by a Royal Society University Research Fellowship for the majority of this work, and acknowledges current financial support from the STFC under grant ST/X000931/1. JCH acknowledges support from the UNAM-PAPIIT grant IG102123 “Laboratorio de Modelos y Datos (LAMOD) para proyectos de Investigación Científica: Censos Astrofísicos”, as well as from SECIHTI (formerly CONAHCYT) grant CBF2023-2024-162 and DGAPA-PAPIIT-UNAM grant IN110325 “Estudios en cosmología inflacionaria, agujeros negros primordiales y energía oscura.”

\appendix
\section{Apparent horizon velocity}\label{App:velocity}

In this appendix, we provide details on the derivation of Eqs.~\eqref{eq:VAH_tilde} and \eqref{eq:alpha}. To this end, we closely follow Ref.~\citep{Helou:2016xyu}. We introduce the Lie derivatives of $\Theta^p = (\Theta^+, \Theta^-)$ along the congruence $k^\mu_q = (k^{\mu (+)}, k^{\mu (-)})$, defined as
\begin{equation}
\mathcal{L}_p \Theta^q = k_p^\mu \partial_\mu \Theta^q = \left(e^{-\phi} \partial_t + p\, e^{-\lambda/2} \partial_A\right) \Theta^q,
\end{equation}
where $p$ and $q$ take values $(+,-)$. The causal nature of an apparent (or cosmological) horizon can be characterized by the parameter
\begin{equation}
    \alpha \equiv 
    \frac{\mathcal{L}_p \Theta^p}{\mathcal{L}_{\mathrm{no}\text{-}p} \Theta^p},
\end{equation}
where evaluation at the horizon is implicit and $\alpha$ is related to the horizon velocity as shown in Eq.~\eqref{eq:VAH_tilde}. In particular, for a PBH apparent horizon where $\Theta^+ \propto (U + \Gamma)$, we obtain
\begin{eqnarray}
    \alpha = \frac{\mathcal{L}_+ \Theta^+}{\mathcal{L}_- \Theta^+} &=& \frac{\left(e^{-\phi} \partial_t + e^{-\lambda/2} \partial_A\right)(U + \Gamma)}{\left(e^{-\phi} \partial_t - e^{-\lambda/2} \partial_A\right)(U + \Gamma)} \nonumber \\
    &=& \frac{e^{-\phi} \partial_t T + e^{-\lambda/2} \partial_A T}{e^{-\phi} \partial_t T - e^{-\lambda/2} \partial_A T},
\end{eqnarray}
where $T \equiv U + \Gamma$. Computing the derivatives, we obtain:
\begin{subequations}
\begin{eqnarray}
    e^{-\phi}\partial_t T &=& -\left[\frac{e^{-\lambda/2}\partial_A P_{\rm pf}}{\rho_{\rm pf}+P_{\rm pf}}(U+\Gamma) + 4\pi R(P_{\rm T}+e^{-\lambda/2}\chi\Pi) \right. \nonumber \\
    && \left. + \frac{1+U^2 -\Gamma^2}{2R}\right], \\
    e^{-\lambda/2}\partial_A T &=& \frac{U+\Gamma}{\Gamma}e^{-\lambda/2} \partial_A U + \frac{1+U^2 - \Gamma^2}{2R} - 4\pi R \rho_{\rm T} \nonumber \\
    && + 4\pi \frac{U}{\Gamma} e^{-\lambda/2} R \chi \Pi.
\end{eqnarray}
\end{subequations}
Using the above expressions and evaluating at the apparent horizon ($U = -\Gamma$), we obtain
\begin{equation}
    \alpha = \frac{4\pi R^2\left(\rho_{\rm T} + P_{\rm T} + 2e^{-\lambda/2}\chi\Pi\right)}{1 - 4\pi R^2\left(\rho_{\rm T} - P_{\rm T}\right)},
\end{equation}
which, once re-expressed in terms of tilde quantities, reduces to Eq.~\eqref{eq:alpha}.

\section{Hamiltonian constraint relation}\label{App:HC}
By differentiating Eq.~\eqref{eq:Gam1} with respect to $A$ and rearranging terms, we obtain:
\begin{equation}
{UU'- 4\pi \rho_{\rm T} R R^{'} + \frac{m_{\rm T}}{R^2} R'  + 4\pi RU \chi \Pi}-  \frac{I}{R^2} R'  -{\Gamma \Gamma'} = 0,
\end{equation}
where, for simplicity, we have again adopted the convention that primes denote derivatives with respect to $A$. Using the relations $\Gamma / R' = e^{-\lambda/2}$ and $(I - m_{\rm T}) / R^2 = (\Gamma^2 - U^2 - 1) / (2R)$, the above expression becomes:
\begin{eqnarray}
&&{UU'- 4\pi \rho_{\rm T} R R^{'} + 4\pi RU \chi \Pi} -\frac{\Gamma^2 R^{'}}{2R} \nonumber\\
&&+\frac{\Gamma^2 R^{'}}{2R}+\frac{R^{'}}{2R}  -{\Gamma \Gamma'} = 0,
\end{eqnarray}
In terms of dimensionless (tilde) variables, this relation reduces to Eq.~\eqref{eq:HC}.


\bibliographystyle{ieeetr}
\bibliography{biblio}

\end{document}